\documentclass[useAMS]{mn2e}
\usepackage{graphicx}

\title[]{}
\author[]{}

\title[Central MONDian Spikes] {Central MONDian spike in spherically symmetric systems} 
\author[X. Hernandez] {X. Hernandez \\Instituto de Astronom\'{\i}a,
  Universidad Nacional Aut\'{o}noma de M\'{e}xico, Apartado Postal
  70--264 C.P. 04510 Ciudad de M\'exico, M\'exico.} 

\date{Released 15 Jan. 2017}

\pagerange{\pageref{firstpage}--\pageref{lastpage}} \pubyear{2015}

\begin{document}

\label{firstpage}

\maketitle

\begin{abstract}
Under a MONDian view, astrophysical systems are expected to follow Newtonian dynamics whenever the local
acceleration is above the critical $a_{0}=1.2 \times 10^{-10} m s^{-2}$, and enter a modified regime for
accelerations below this critical value. Indeed, the dark matter phenomenology on galactic and
subgalactic scales appears always, and only,
at low accelerations. It is standard to find the $a<a_{0}$ regime towards the low density outskirts
of astronomical systems, where under a Newtonian interpretation, dark matter becomes conspicuous. Thus, it is
standard to find, and to think, of the dense central regions of observed systems as purely Newtonian.
However, under spherical symmetry in the MONDian as in the Newtonian case, the local acceleration will tend
to zero as one approaches the very centre of a mass distribution. It is clear that for spherically symmetric
systems, an inner $a<a_{0}$ region will necessarily appear interior to a critical radius which will depend on
the details of the density profile in question. Here we calculate analytically such a critical radius for a
constant density core, and numerically for a cored isothermal profile. Under a Newtonian interpretation,
such a central MONDian region will be interpreted as extra mass, analogous to the controversial black holes
sometimes inferred to lie at the centres of globular clusters, despite an absence of nuclear activity detected
to date. We calculate this effect and give predictions for the ``central black hole'' mass to be expected
under Newtonian interpretations of low density Galactic globular clusters.

\end{abstract}


\section{Introduction} \label{intro}

The continual null detection of any dark matter particles, in particular over the last year, LHC results 
eliminating simple super-symmetric candidates (CMS collaboration 2016), the astrophysical searches for dark matter
annihilation signals being fully consistent with zero dark matter signal (e.g. Fermi-LAT and DES collaborations 2016
searching for such a signal in local dwarf galaxies reporting results consistent with expected backgrounds), and
various recent direct detection experiments returning only ever stricter exclusion limits (e.g. Yang et al. 2016
reporting no dark matter signal from the PANDAX-II experiment, ruling out previous claims, and Szydagis et al. 2016
for the LUX and LZ collaborations reporting also no dark matter signal), encourage the sustained exploration of
alternative explanations for the gravitational anomalies appearing in the low acceleration regime, and generally
ascribed to the presence of dark matter.

Further, recent theoretical developments have shown novel possible
fundamental physical origins for a change in regime for gravity when reaching the low acceleration region of MOND,
e.g. the emergent gravity ideas of Verlinde (2016) {  (but see also Lelli et al. 2017, Hees et al. 2017,
Tortora et al. 2017)}, or the covariant models including torsion of Barrientos \& Mendoza
(2016). Empirically, Durazo et al. (2017) analysing pressure supported systems from globular clusters to elliptical
galaxies, and McGaugh et al. (2016) and Desmond (2017) studying spiral rotation curves, all find the ratio between
inferred dark matter content and observed baryonic mass to be consistent with MONDian expectations across a wide
range of galaxy types and masses.

In light of the above, it appears reasonable to continue the study of MONDian gravity and its predictions.
In this paper we present a novel effect, the expectation of a modified gravity region towards the centre
of astrophysical systems, and perform a first exploration of its consequences. Under MONDian gravity the usual
Newtonian physics are recovered in the high acceleration regime, while whenever local acceleration falls below
the critical $a_{0}=1.2 \times 10^{-10} m s^{-2}$ value, the gravitational force shifts towards $(G M a_{0})^{1/2}/R$,
for test particles orbiting at a radius $R$ from a point mass $M$. This low acceleration regime typically
occurs towards the outskirts of astrophysical systems, where lower matter densities naturally lead to low
accelerations. It becomes clear however, that given the validity of Newton's theorems for spherical mass
distributions in both Newtonian and MONDian gravity (e.g. Mendoza et al. 2011), the local acceleration will tend to
zero for $R \rightarrow 0$. Necessarily, this will lead to an inner MONDian region on crossing inwards
of a threshold radius interior to which $a<a_{0}$, as $a \rightarrow 0$ in going to the actual centre of
an astrophysical system.

We present a first development of this effect for constant stellar density cores through a simple analytic
development to show the physical scalings and orders of magnitude expected for the size of this
inner MONDian region, as well as its expected force amplitude, as a function of the density of such a region.
We show that if dynamical tracers within this enhanced gravity region are interpreted under Newtonian
assumptions, an extra mass component would be inferred, perhaps akin to the intermediate mass black holes
reported by some authors (e.g. Ibata et al. 2009, Feldmeier et al. 2013, Kamann et al. 2016) in the centres
of Galactic globular clusters, and which have failed to appear under direct searches (e.g. Maccarone \& Servillat
2010, Lu \& Kong 2011 and Strader et al. 2012 report only upper limits on GC black holes through accretion activity
diagnostics). A numerical treatment for cored isothermal stellar distributions, intended merely as accurate
phenomenological descriptions of the observed density profiles of globular clusters, is also performed, with
parameters representative of Galactic globular clusters, in order to explore and give some predictions regarding
the gravitational anomaly expected under MONDian gravity towards the centres of these systems.
Such region should range from somewhat over the inner $10 pc$ to less than $1 pc$, in going from systems
with central densities spanning from a few tens to a few thousands of $M_{\odot} pc^{-3}$.

In section 2 we derive analytically the details of this effect for constant density cores, showing the physical
scalings which result. In section 3 we extend the results numerically for cored isothermal stellar profiles,
as representative models of the actual stellar density profiles of observed globular clusters. In this section
we also compare with recent studies addressing the controversy of intermediate mass black holes in globular
clusters, and give predictions for such an effect in low density globular clusters. Our concluding remarks appear
in section 4.

\section{Central MONDian spikes}

Within a MONDian gravity scheme one expects the force felt by a test particle orbiting a total mass $M$
to be well described by the Newtonian expression of $F_{N}(R)=-GM/R^{2}$ whenever the acceleration is above $a_{0}$,
and to follow the MONDian expression of $F_{M}(R)=-(G M a_{0})^{1/2}/R$ in the low acceleration regime. The transition
between regimes appears at the point where $F_{N}(R)=F_{M}(R)$, yielding a characteristic MOND radius
$R_{M}= (G M/a_{0})^{1/2}$ e.g. Milgrom (1984), Mendoza et al. (2011). The condition $a<a_{0}$ for the modified regime
yielding directly $R>R_{M}$, the MONDian region appears at distances larger than the characteristic $R_{M}$.

Given the validity of Newton's theorems for spherically symmetric mass distributions also under the MONDian regime
(e.g. Mendoza et al. 2011), the above expressions can be generalised directly by substituting $M \rightarrow M(R)$.
If we now assume a constant density core, as appropriate for a wide range of astrophysical systems, e.g. the central
regions of the King halos generally fitted to observed surface brightness profiles of Galactic globular clusters,
we can take $M(R)=4 \pi \rho_{0} R^{3}/3$ to yield:

\begin{equation}
R_{M}=\left( \frac{4 \pi G \rho_{0} R^{3}}{3 a_{0}} \right)^{1/2},
\end{equation}

\noindent where $\rho_{0}$ is the density of the core region in question. As the condition $a<a_{0}$ remains
$R>R_{M}$, and as $R_{M}$ scales with $R^{3/2}$, it is clear that $R_{M}$ will fall to zero towards the centre
at a faster rate than the radial coordinate, yielding necessarily an inner critical radius interior to which
a central MONDian region is to be expected. This critical radius, $R_{c}$, can be found by equating the radial
coordinate to $R_{M}$, which gives:

\begin{equation}
R_{c}=\frac{3 a_{0}}{4 \pi G \rho_{0}}.  
\end{equation}

\noindent the previous relation in astronomical units reads:

\begin{equation}
R_{c} =211.3 \left ( \frac{M_{\odot} pc^{-3}}{ \rho_{0}} \right)  pc.   
\end{equation}

For a constant density core, interior to the above radius dynamics will show a transition towards a MONDian
regime. We see that in a globular cluster with a typical central density of $100 M_{\odot} pc^{-3}$ only the
central $2 pc$ will lie within the $a<a_{0}$, $R>R_{M}$ inner modified regime. While in going to high density globular
clusters, with central densities an order of magnitude higher, the central MONDian region will be limited
to a tiny $0.2 pc$ region. For low density systems, the modified regime could easily extend to the central
$10 pc$ or more. The development leading to equation (1) can be repeated for a stellar density profile
having a central scaling $\rho (R) \propto R^{n}$ where the condition for $R_{M}$ to fall to zero towards
$R \rightarrow 0$ faster than the radial coordinate becomes $n>-1$. Thus, we see that for all stellar
profiles having a central cusp shallower than $R^{-1}$, an internal low acceleration $a<a_{0}$ region
will exist. {  A clear precedent for this effect can be found in the study by Ciotti et al. (2006),
where in their figure (1) a ``dark matter'' spike appears for the cases of central baryonic density
distributions shallower than  $R^{-1}$, as shown above.}

In the above we have ignored the details of any MOND transition function and assumed an abrupt shift from
the Newtonian to the MONDian regimes. Such details, in MOND as such, are not well constrained beyond the
requirement of a fairly abrupt transition e.g. Milgrom (2014). In
Hernandez \& Jimenez (2012) we showed that under a MONDian modelling of Galactic globular cluster structure, that
when considering MONDian force laws at the Newtonian level, the transition function is required to be
very sudden, as also implied in order to avoid any conflict with solar system
dynamics.

We can now calculate what extra central mass will be inferred under Newtonian dynamics from dynamical tracers
within $R_{c}$ by equating the actual force at a given radius, to the corresponding Newtonian expression,
to which we add a hypothetically central potential:

\begin{equation}
\frac{[G M(R) a_{0}]^{1/2}}{R} = \frac{G M(R)}{R^{2}} +\frac{G M_{BH}}{R^{2}}.
\end{equation}

If we now go to the constant density core of a baryonic mass distribution, e.g. the central regions of
Galactic globular clusters, $M(R)=4 \pi \rho_{0} R^{3}/3$ and we can solve for $M_{BH}$ from the above expression
to yield:

\begin{equation}
  M_{BH}=\left( \frac{4 \pi \rho_{0} R^{5}}{3}  \right)^{1/2} \left[ \left( \frac{a_{0}}{G}\right)^{1/2} -
    \left(  \frac{4 \pi \rho_{0} R}{3}   \right)^{1/2}      \right]
\end{equation}

We see that for large densities and radii the above equation becomes negative, whenever $R<R_{M}$
c.f. equation (2), the Newtonian term dominates and the equation is not valid, as no MONDian regime appears.
Also, notice that the inferred black hole mass under a Newtonian interpretation is a function of the radial
distance at which the tracers being used for the inference are found, i.e., no consistent, unique inferred black
hole mass will appear when treating tracers found over a range of radial distances. Of course, if tracers within
the central MONDian spike are analysed under a Newtonian assumption, some extra mass will necessarily be required,
and if one then fits a model with such a black hole mass as a free parameter, statistically, some preferred
value will appear. 

We can estimate the maximum black hole mass which under a Newtonian interpretation could appear as a result of the
central MONDian region, by setting the radial derivative of the previous equation equal to zero and finding the
radius at which a maximum black hole mass appears, $R_{MX}$ given by:

\begin{equation}
R_{MX}=\left( \frac{25}{36} \right) R_{c},  
\end{equation}  

\noindent i.e., slightly inside of the outer edge of the MONDian central region one finds the radius at which
tracers will, under a Newtonian interpretation, require a maximal black hole mass to fit dynamics. This maximum
black hole mass will now be given by:

\begin{equation}
  M_{BH}(R_{MX}) = \left( \frac{a_{0}^{3}}{G^{3}\rho_{0}^{2}} \right) \left( \frac{4 \pi}{3} \right)^{1/2}
  \left( \frac{25}{48 \pi} \right)^{5/2} \left[ 1- \left( \frac{25 \pi}{36}\right)^{1/2} \right],
\end{equation}

\begin{equation}
M_{BH}(R_{MX}) =  3.82\times 10^{-3} \left( \frac{a_{0}^{3}}{G^{3}\rho_{0}^{2}} \right)  M_{\odot},
\end{equation}

\noindent which going to astronomical units yields:

\begin{equation}
M_{BM}(R_{MX}) = 2.65\times 10^{6} \left( \frac{M_{\odot} pc^{-3}}{\rho_{0}} \right)^{2} M_{\odot}.  
\end{equation}

Hence, for central globular cluster densities of below $51.3 M_{\odot} pc^{-3}$, in the range of what is
inferred for some well studied Galactic globular clusters (e.g. Harris 1996 and updates thereof), central
black hole masses of more than $1000 M_{\odot}$ will be required to understand the dynamics of stellar
tracers within a constant density region, under Newtonian assumptions. For higher central globular cluster
densities, the maximum required black hole mass drops as shown above, with the square of the core density.

\section{Predictions for Galactic Globular Clusters}

In the previous section we have given analytic estimates for the effect being presented, under the assumption
of a core region of rigorously constant stellar density. In reality, even within the core region of King
halos, as commonly used to describe observed globular clusters, a slight radial drop in density occurs. Since we
are clearly dealing with an effect which is very centrally located, and since all King halos tend strongly
within their core regions to a cored isothermal solution (see e.g. fig 4-9 in Binney \& Tremaine 1987), we can
accurately estimate the effect of the slight curvature in the density profile within the core region of Galactic
globular clusters, without reference to particular central potential values, by modelling the situation through
a cored isothermal stellar density profile, intended solely as an empirical description of the density profile
of observed globular clusters.

We begin by calculating cored isothermal profiles, by solving numerically the equation:

\begin{equation}
\sigma^{2} \frac {d\rho}{dR} =- \rho \frac{G M(R)}{R^{2}},
\end{equation}

\noindent under the conditions $\rho(0)=\rho_{0}$, $d\rho/dR =0$ at the origin and taking a fixed velocity
dispersion of $\sigma = 5 km/s$, a value representative of what is observed in Galactic globular clusters, e.g.
Harris (1996), Scarpa et al. (2011). The central stellar densities of these profiles were chosen to span a range
of values bracketing that which is inferred for Galactic globular clusters, we used 50, 100, 300, 1000 and 3000
$M_{\odot}pc^{-3}$.

\begin{figure}
\includegraphics[width=9.0cm,height=7.0cm]{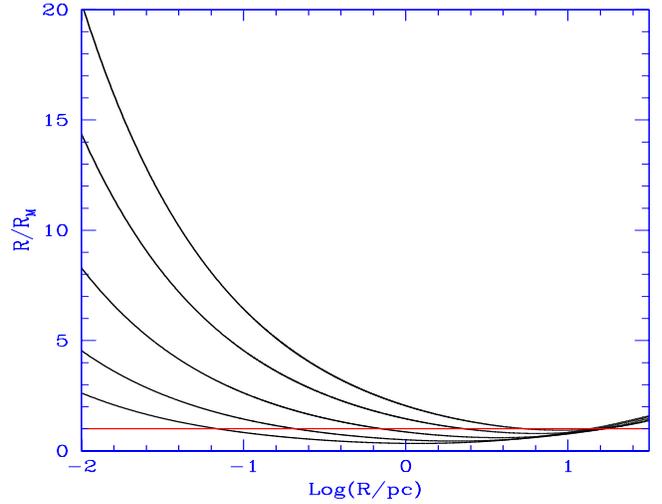}
\caption{The figure gives the ratio of the radial coordinate to the local MOND radius, $R/R_{M}$, as a function
of the logarithm of the radial distance in pc, for cored isothermal halos characterised by a velocity dispersion
of $\sigma=5 km/s$ and central density values of 50, 100, 300, 1000 and 3000 $M_{\odot}pc^{-3}$, top to bottom
respectively. The horizontal line gives $R=R_{M}$, with values above indicative of a central MONDian spike.
}
\end{figure}

Once the full density profiles and corresponding mass profiles $M(R)$ were calculated, we evaluated the ratio
between the radial coordinate and the local MOND radius of $(G M(R)/a_{0})^{1/2}$. The result is shown in figure (1),
where said ratio appears as a function of the logarithm of the radial distance for the cored isothermal
halos mentioned above, top to bottom for decreasing values of the central density. The horizontal line in the
figure gives $R=R_{M}$, with the central MONDian region appearing interior to the first crossing of the
corresponding curves and this line. These crossing points are always within a few percent of the analytic
estimates for rigorously constant stellar density cores of equation (2).

We now solve for the Newtonian central black hole mass required to reproduce the actual MONDian radial force
within the internal $R>R_{M}$ region, when added to the Newtonian force produced by the actual stellar density
profile present, i.e., we solve for $M_{BH}$ from equation (4), using this time the full cored isothermal
$M(R)$ profiles described above. The results are shown in figure (2), where in a log-log plot we give the
central black hole mass which would be inferred under a Newtonian modelling of dynamical tracers observed
within the halos described above. It is interesting that the actual maximum values for these masses, and the
radii at which these occur, are never more than a few percent off from the rigorously constant stellar
density core results of the previous section. As expected from the analytic results, the maximum
black hole mass appears just inside the onset of the central MONDian spike, the $M_{BH}(R)$ functions
in figure (2) drop very abruptly shortly after reaching their maxima.

\begin{figure}
\includegraphics[width=9.0cm,height=7.0cm]{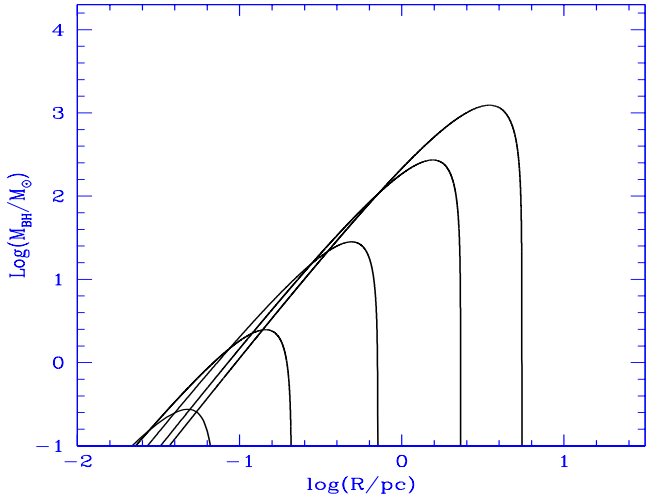}
\caption{Assuming dynamical tracers at each radius were interpreted under a Newtonian scheme, the figure gives the
resulting central extra mass which would be inferred, in a log-log plot, for cored isothermal halos characterised by
a velocity dispersion of $\sigma=5 km/s$ and central density values of 50, 100, 300, 1000 and 3000 $M_{\odot}pc^{-3}$,
top to bottom respectively. Maximum values and corresponding radii are within a few percent of the results
of equation (9) for flat cores of the same densities.}
\end{figure}

Going back to figure (1), we see that the inner MONDian region extends outwards as the central density assumed is
reduced, in consistency with equation (3). In going to progressively lower densities, the inner $a<a_{0}$
region extends to cover a growing percentage of the cluster extent, until a density threshold is reached at
which the inner MONDian region extends into the typical $a<a_{0}$ outer low acceleration one, and the globular
cluster appears entirely MONDian. Indeed, such is the case of NGC 288, a very low desnsity system  fully within
the low acceleration regime. As expected under MONDian gravity, Hernandez et al. (2017) recently found that this
cluster shows no drop in its velocity dispersion profile, being fully isothermal and fully consistent in both
velocity dispersion and projected surface density profiles with MONDian dynamical models.

{  As mentioned previously, in the preceding development we have assumed that the force law which applies
is the one which corresponds to the $a<<a_{0}$ limit of the deep MOND regime, that is, the l.h.s. of equation (4)
considers a purely MOND-limit force law. This would be accurate in the case of an abrupt transition
between the Newtonian and the modified regime, for example, if this transition proved to be of quantum origin.
If however, a more gradual transition between the two limit regimes applies, as is generally assumed
through the $\mu(a/a_{0})$ transition functions of MOND (e.g. Famaey \& McGaugh 2012), the l.h.s.
of equation (4) would be modified to include a more complicated expression accommodating a gradual
transition to the standard Newtonian expression in the $a>>a_{0}$ limit. As can be seen in figure (1),
the $a_{0}$ transition point is approached quite gradually for the cored isothermal density distribution
profiles considered, the case for constant density cores is quite similar. This last means that for
gradual transition functions, the resulting MONDian spike could be quite different than for the limit
case shown in figures (1) and (2).

In Bekenstein (2004) it is shown that for the relativistic extension of TeVeS, the function which mediates
the transition between the MOND and the Newtonian regimes for a MOND description of $a \mu(X)=GM/R^{2}$,
where $X=a/a_{0}$, is:

$$
\mu{X}=\frac{(1+4X)^{1/2}-1}{(1+4X)^{1/2}+1},
$$

\noindent the so called 'simple' MOND transition function is:
$$
\mu(X)=\frac{X}{X+1},
$$
\noindent while the 'n-family' of transition functions is:

$$
\mu{X}=\frac{X}{(X^{n}+1)^{1/n}}.
$$

Taking $n=2$ in the last expression yields what is generally referred to as the standard
$\mu$ function. All of the above clearly have the required limits of $\mu(X) \rightarrow 1$
for $X>>1$ and  $\mu(X) \rightarrow X$ for $X<<1$. Figure (3) is analogous to figure (2),
but shows in all cases the results for a central density value of $300 M_{\odot} pc^{-3}$,
for different assumed $\mu(X)$ functions, top to bottom, the Bekenstein transition function,
the simple MOND $\mu(X)$, the 'standard' MOND function, followed by $n=4, 8, 16$ and $32$ of
the 'n-family'. The lowermost curve is the $a<<a_{0}$ limit shown in figure (2)

We see that for the cases of very soft transition functions, the MONDian modification of the mildly
Newtonian regime for accelerations slightly above $a_{0}$, is significant and results in extremely
large gravitational anomalies which from the point of view of Newtonian dynamics would be interpreted
as extra matter components. In fact, these soft transition functions are already ruled out by
solar system consistency checks, where the absence of measurable discrepancies with regards to
Newtonian physics eliminates $\mu(X)$ functions which introduce even small modified gravity
effects at solar system acceleration scales, e.g. Mendoza et al. (2011). Even low n-family
transition functions are hard to accommodate, e.g., Hernandez \& Jimenez (2012) showed that
dynamical models for Galactic globular clusters constrained to fit both surface brightness and
velocity dispersion profiles required very abrupt $\mu(X)$ functions. As progressively
higher $n$ values are taken, it is clear that the transition function becomes more and more abrupt,
with the results tending to the limiting MOND case of equation (4) and shown in figure (2).
Thus, we propose that detailed studies of the radial range over which this central MONDian
spike can be detected (or alternatively intermediate mass black holes in the absence of any
high energy activity signature under Newtonian assumptions), can serve to constrain the abruptness
of the MOND transition function.}

{  As discussed above, the systems under consideration hover close to the transition $a=a_{0}$
point, so that a return to a scaled Newtonian gravity in the deep $a<<a_{0}$ regime, as happens when an
$\epsilon_{0}$ term is included in the transition function (Famaey \& McGaugh 2012) would have negligible
consequences.}

\begin{figure}
\includegraphics[width=9.0cm,height=7.0cm]{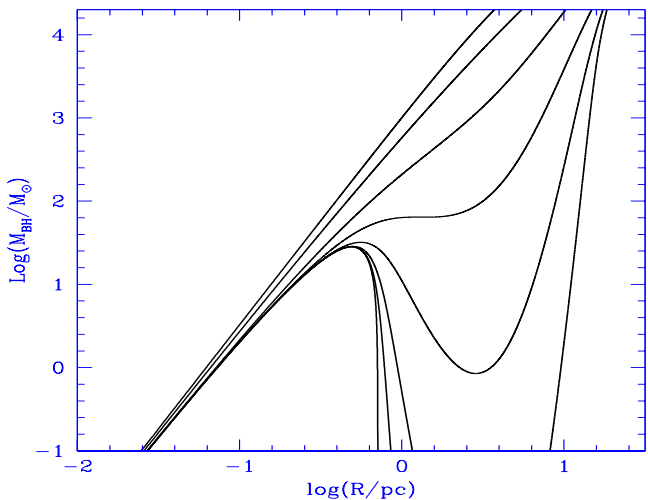}
\caption{This figure is analogous to figure (2), but shows in all cases the results for a central density
value of $300 M_{\odot} pc^{-3}$, for different assumed $\mu(X)$ functions, top to bottom, the Bekenstein
transition function, the simple MOND $\mu(X)$, the 'standard' MOND function, followed by $n=4, 8, 16$ and $32$.
The lowermost curve is the $a<<a_{0}$ limit shown in figure (2)}
\end{figure}

{  A further caveat to what we present here is that under MOND as such, some Galactic globular
clusters would be expected to behave as essentially Newtonian systems throughout, even if their internal
accelerations are lower than $a_{0}$. This a a consequence of the so called external field effect, where
if the external gravitational acceleration produced by a larger system -in this case the Milky Way- is
actually larger than $a_{0}$, the internal dynamics would revert to a scaled Newtonian character e.g.
Famaey \& McGaugh (2012). It is not known if this effect appears at all under the various covariant
plausible theoretical origins of the MOND phenomenology proposed, e.g. Barrientos \& Mendoza (2016)
or Verlinde (2016) and indeed, systems showing evidence of MONDian phenomenology in cases where the EFE
would predict essentially a return to Newtonian dynamics have been reported, e.g. kinematics of wide
binaries in the solar neighbourhood (Hernandez et al. (2012b) or Galactic globular clusters asymptotically
following the deep MOND expected "Tully-Fisher" scaling between their total baryonic content and their
velocity dispersion profiles (Hernandez et al. (2013). Still, if the external field effect of MOND as such
proves to be a reality, the analysis presented here would hold only for the more distant Galactic globular
clusters which lie at locations where their orbital accelerations about the Milky Way are actually lower
than $a_{0}$, e.g. Sanders (2012) for NGC 2419. Indeed, recent proper motion determinations by Massari et
al. (2017) have determined this particular cluster has an orbit which results in a Galactocentric radial
variation of between 53 and 98 kpc.}

Over the past years, a number of authors have claimed inferences of black holes at the centres of Galactic
globular clusters, e.g. Ibata et al. (2009), Feldmeier et al. (2013) and Kamann et al. (2016). Black hole
claims in globular clusters remain controversial, as direct searches looking for any central nuclear activity
tend to return only upper limits on the putative black hole masses, e.g.  Maccarone \& Servillat
(2010), Lu \& Kong (2011) and Strader et al. (2012). Indeed, Baumgardt (2017) through an extensive Newtonian
grid of N-body simulations calibrated to specific globular clusters where black hole masses have been claimed,
finds that no such black holes are needed, or are even compatible with the observed joint surface density and
velocity dispersion observations, above about $1000 M_{\odot}$, much below the typical claimed black hole masses.
The only exception to the above is the case of $\omega$ Cen, where  Baumgardt (2017) does find evidence for
a $\sim 40,000 M_{\odot}$ central black hole.

In the context of our study, given the central densities of the
globular clusters for which central black holes have been inferred, of order $10^{4} M_{\odot} pc^{-3}$ and
above (Harris 1996), the recent results of Baumgardt (2017) are consistent, as we would expect only very small
``black hole'' masses to be required under Newtonian modelling of the central MONDian region of these clusters
of less than $1 M_{\odot}$ c.f. equation (9) under abrupt $\mu(X)$ transition functions. It is of course not
impossible that some of these globular clusters, due to dynamical evolution processes and their high stellar
densities, might actually contain central black holes, as might probably be the case of $\omega$ Cen. Indeed,
the added central potential produced by the inner MONDian region could play a part in enhancing the formation
efficiency of central black holes in globular clusters.

As changes in the transition function away from the abrupt limit of equation (4) yield always increasing
``black hole masses'', the effect presented here is falsifiable, the lack of any inferred black hole
(under Newtonian assumptions) at the level of the predictions of equation (9) for suitably located dynamical
tracers, would clearly invalidate the hypothesis of the analysis presented, and challenge the MONDian view point.
Thus, a prediction remains for gravitational anomalies in the centres of lower central density globular clusters.
A more solid understanding of the effect of this central MONDian region in the context of dynamical evolution,
mass segregation and N-body relaxation of globular clusters lies beyond the scope of this first presentation of
the effect, but would certainly be desirable to more fully asses its consequences.

\section{Final remarks}

We have shown that for stellar density profiles having a central cusp shallower than $\rho \propto R^{-1}$,
an inner low acceleration region will exist where $a<a_{0}$. For the particular case of a constant density
core of density $\rho_{0}$, the extent of this inner region will be of $211.3 (\rho_{0}/M_{\odot} pc^{-3}) pc$.

Under MONDian gravity, this implies that within this inner region, dynamical tracers will behave as if under
the gravitational influence of an extra mass component, when interpreted under Newtonian assumptions.

If this extra potential under Newtonian interpretations is ascribed to a central black hole, its mass will
depend on the radial range over which dynamical tracers are used, with a maximum inferred black hole
mass of $M_{BM}(R_{MX}) = 2.65\times 10^{6} \left( \frac{M_{\odot} pc^{-3}}{\rho_{0}} \right)^{2} M_{\odot}.$

The previous expression becomes hence a prediction for Newtonian inferences from volume resolved dynamics
of $\rho_{0} \sim 100 M_{\odot} pc^{-3}$ Galactic globular clusters.

{  As the $a_{0}$ threshold is approached gradually, the effect described is very sensitive to the details
of the MONDian transition function used, this presents the possibility of tightly constraining this transition
function through detailed measurements of the extra central ``dark matter'' in galactic globular clusters.}

\section*{acknowledgements}

This work was supported in part by DGAPA-UNAM PAPIIT IN-104517 and CONACyT. I thank an annonymous referee for
his/her careful reading of the first version of this manuscript, this final version was enriched by
valuable constructive criticism found in her/his report.


\begin{thebibliography}{99}

\bibitem{} Barrientos E., Mendoza S., 2016, arXiv:1612.07970

\bibitem{} Baumgardt H., 2017, MNRAS, 464, 2174

\bibitem{} Bekenstein, J. D. 2004, Phys. Rev. D, 70, 083509  
  
\bibitem{} Binney J., Tremaine S., 1987, Galactic Dynamics, Princeton University Press, Princeton.

\bibitem{} Ciotti L., Londrillo P., Nipoti C., 2006, ApJ, 640, 741
  
\bibitem{} CMS collaboration, 2016, arXiv:1609.02507

\bibitem{} Desmond H., 2017, MNRAS, 464, 4160  

\bibitem{} Durazo R., Hernandez X., Cervantes Sodi B., Sanchez, S. F. 2017, ApJ, 837, 179

\bibitem{} Famaey B., McGaugh S. S., 2012, LRR, 15, 10
  
\bibitem{} Feldmeier A. et al., 2013, A\&A, 554, A63
  
\bibitem{} Fermi-LAT collaboration, 2016, arXiv:1611.03184, ApJ in press

\bibitem{} Harris, W.E. 1996, AJ, 112, 1487

\bibitem{} Hees A., Famaey B., Bertone G., 2017, arXiv:170204358
  
\bibitem{} Hernandez X., Jimenez M. A., 2012, ApJ, 750, 9

\bibitem{} Hernandez X., Jimenez M. A., Allen, C., 2012, EPJC, 72, 1884

\bibitem{} Hernandez X., Jimenez M. A., Allen C. 2013b, MNRAS, 428, 3196
  
\bibitem{} Hernandez X., Cortes R. A. M., Scarpa R., 2017, MNRAS, 464, 2930

\bibitem{} Ibata R. et al., 2009, ApJ, 699, L169

\bibitem{} Kamann S. et al., 2016, A\&A, 588, A149

\bibitem{} Lelli F., McGaugh S. S., Schombert J. M., 2017, MNRAS, 468, L68

\bibitem{} Lu T., Kong A. K. H., 2011, AJ, 729, L25
  
\bibitem{} Maccarone T. J., Servillat M., 2010, MNRAS, 408, 2511

\bibitem{} Massari D., Posti L., Helmi A., Fiorentino G., Tolstoy E. 2017, A\&A, 598, L9
   
\bibitem{} McGaugh S.S., Lelli F., Schombert J., 2016, Phys. Rev. Lett. 117, 20110

\bibitem{} Mendoza S., Hernandez X., Hidalgo J. C., Bernal T., 2011, MNRAS, 411, 226

\bibitem{} Milgrom M., 1984, ApJ, 287, 571

\bibitem{} Milgrom M., 2014, MNRAS, 437, 2531

\bibitem{} Sanders, R. H. 2012, MNRAS, 419, L6  

\bibitem{} Scarpa R., Marconi G., Carraro G., Falomo R., Villanova S., 2011, A\&A, 525, 148

\bibitem{} Strader J., Chomiuk L., Maccarone T. J., Miller-Jones J. C. A., Seth A. C.,
Heinke C. O., Sivakoff G. R., 2012, ApJ, 750, L27
  
\bibitem{} Szydagis et al. 2016, arXiv:1611.05525

\bibitem{} Tortora C., Koopmans L. V. E., Napolitano N. R., 2017, arXiv:170208865

\bibitem{} Yang et al. 2016, arXiv:1612.01223

\end{thebibliography}
\end{document}